\documentclass[11pt]{article}
\usepackage{graphics}
\usepackage{epsfig}

\evensidemargin=\oddsidemargin
\def\ingnore{\ifnum 1=0}

\def\l@subsection{\@dottedtocline{2}{3.5em}{2.0em}}
\def\l@subsubsection{\@dottedtocline{3}{5.5em}{2.0em}}

\begin{document}

\title{Instabilities in Chains Coupled by Two-Body Interactions}
\author{F.Vistulo de Abreu, B. Dou\c cot \\ 
Laboratoire de Physique Th\'eorique et Hautes Energies,Jussieu, 75252 Paris
CEDEX 05\\
\emph{and}\\
Centre de Recherches sur les Tr\`es Basses Temperatures\\
Grenoble BP 166X CEDEX}
\maketitle

\begin{abstract}
We derive a general set of Poor Man's scaling equations and analyze the
stability of the Luttinger state in a system composed of a finite number N
of one dimensional spinless fermionic chains, coupled through a general two
body interaction. The effect of processes with momentum transfer parallel to
the Fermi surface in destroying massless states is investigated. It will be
shown that there are two processes competing: one in which two electrons
exchange chains and the other in which they jump into a same chain. When
periodic boundary conditions in the transverse direction are taken into
account this competition leads always to massive states (except in
hyperplanes of the phase diagram), a well known example being the
generalized sine-Gordon model. If instead open boundary conditions are
taken, massless states are possible but due to this competition the system
is placed near instabilities. We argue that this kind of analysis has
relevance for understanding the instabilities of 2D fermionic systems.

Submitted to Europhysics Letters: 1st version 21 Juin 1996; revised version
2nd October 1996
\end{abstract}

The many unexplained features of high-Tc superconductors have motivated an
intense study of low dimensional interacting fermions\cite{Houghton93}. A
renormalization group (RG) approach then developed allowed a reformulation
of the Fermi liquid concept\cite{Shankar94}. However this analysis seems to
rely heavily on the isotropic nature of the Fermi surface(FS). Indeed, for a
two-dimensional(2D) circular FS one can neglect, to some extent, the effect
of scattering processes with momentum transfer tangential to the Fermi
surface. This is because the FS topology highly restricts contributions to
the zero sound channel by a factor of $\left( \Lambda /k_F\right) ^2$ (where 
$\Lambda $ is the momentum cutoff used in the RG procedure), whereas the
Cooper channel receives contributions proportional to $\Lambda /k_F.$ Thus
we can neglect the zero sound contributions to the one loop renormalization
of the four fermion vertex and then we are left with a simple differential
equation that diverges \emph{only} if attractive interactions are present,
indicating the appearance of a superconducting instability. However for a FS
with large flat portions (nested), like in the case of a square FS, this
picture may not apply. Then the scattering processes with momentum transfers
parallel to the FS and involving electrons on parallel sides of the FS, give
supplementary contributions and may play an important role in destabilizing
massless regimes. To understand better the effect of these contributions, we
suggest, in a first reasonable approximation, the study of a model of 1D
chains coupled through a general two body interaction.

It is worth noting that this kind of model has quite a large range of
applications, for it can be related (even if approximately) to 1D spin
chains through the Jordan-Wigner transformation and to
systems of coupled Hubbard chains\cite{Schulz86}. Even though, these are
particular cases of our model since for these models the chains are coupled
through density-density interactions, and the densities are operators
defined on each chain. Consequently the number of electrons on each chain is
conserved. By contrast, this is not a requirement for our model, which has
this new degree of complexity. In particular, some commonly used techniques,
like 1D bosonization, find inconsistencies if one takes account of the
fermionic nature of the field operators correctly, through the Haldane
ladder operators\cite{Vistulo96,Haldane81,Yoshioka95}. For this
reason we derive a general set of RG differential equations based on the
Poor Man's Scaling (PMS) method, thus avoiding the use of the bosonization
tools whenever they lead to incongruities.

Consider an array of N spinless fermionic chains, aligned along $x$. As in
the 1D case we separate the field operator in slow and fast components, $%
\psi (\mathbf{x})=1/\sqrt{N}\sum_{p=\pm }\sum_{n=1}^N\exp (i\,p\,\mathbf{k}%
_n\cdot \mathbf{x)}\psi _{p,n}(\mathbf{x})$, where $p\,\mathbf{k}_n$ is the
momentum Fermi surface point for the chain $n$ and branch $p$. We take all $%
\left( \mathbf{k}_n\right) _x$ equal to $k_F$, a linearized dispersion
relation near the FS, and neglect any Fermi velocity modulation. A general
density-density interaction can be written as: 
\begin{equation}
\mathcal{H}_G=\frac 1{NL}\sum_{i,j,\delta }\sum_q\sum_{k,k^{\prime
}}G_\delta ^{~}(i-\frac \delta 2,j+\frac \delta 2)\ \ R_{i+\delta
/2,k+q}^{+}\ \ L_{j-\delta /2,k^{\prime }-q}^{+}\ \ L_{j+\delta /2,k^{\prime
}}\ \ R_{i-\delta /2,k}  \label{HG}
\end{equation}
where here $L_{j,k}^{+}$ is the creation operator for a left mover on chain
j and with momentum $k$ along the chain. If we take periodic boundary
conditions (PBC) along $y,$ then the couplings depend only on $j-i\equiv
\Delta ;\,G_\delta (I,J)\rightarrow G_\delta (\Delta ),$ with $J=j+\delta
/2,I=i-\delta /2$. From (\ref{HG}) we can see that due to the inversion
symmetry (i$\leftrightarrow $j), hermiticity and boundary conditions, these
couplings must verify the conditions: 
\begin{equation}
\begin{tabular}{lll}
$G_\delta (I,J)=G_{-\delta }(J,I)=G_{-\delta }(I+\delta ,J-\delta
)=G_{-\delta }(N+1-I,N+1-J)$ & \  & (no PBC) \\ 
$G_\delta (\Delta )=G_{-\delta }(\Delta )=G_\delta (-\Delta )=G_{-\delta
}(-\Delta )=G_{\delta \pm N}\left( \Delta \right) =G_\delta \left( \Delta
\pm N\right) $ & \  & (PBC)
\end{tabular}
\end{equation}

Note that the chain index can also be regarded as a spin index. In this way
we can establish a dictionary between the usual g-ology couplings\cite
{Solyom79} and the G's, for a model with 2 coupled chains: 
\begin{equation}
\begin{tabular}{llll}
{\normalsize $G_0(1)/2\equiv g_{2\perp }$} & {\normalsize $G_1(0)/2\equiv
-g_{1\perp }$} & {\normalsize $G_0(0)/2\equiv g_{2||}-g_{1||}$} & 
{\normalsize $G_1(1)/2\equiv 0$}
\end{tabular}
\label{dic}
\end{equation}

Notice that according to the formulation (\ref{HG}), the backscattering
processes are now seen as forward scatterings with spin-flips, and thus
within each chain there are no backscatterings (see figure 1a ). This
formulation will allow us to obtain a compact set of RG equations, which is
particularly useful in treating the N-chain problem.

The PMS equations are derived by requiring invariance of the vertex
calculated in a one loop expansion. There are only two diagrams giving
logarithmic contributions, both with the same magnitude (${\rho _0}/2$ $%
[ln(\omega /D)-{i\pi }/2]$ ) but having opposite signs. Here $\rho _0$ is
the density of states for one spin direction. For convenience we will
redefine from now on the couplings as the adimensional quantities $G_\delta
(\Delta )\rightarrow {G_\delta (\Delta )}/{\pi v_F}$. The RG flow equations
for models with and without PBC are then: 
\begin{eqnarray}
\frac{\partial G_\delta (\Delta )}{\partial \ln D} &=&\frac
1{2N}\sum_{\alpha =0}^{N-1}\ [G_\alpha (\Delta +\delta -\alpha )G_{\delta
-\alpha }(\Delta -\alpha )-G_{\delta -\alpha }(\Delta )G_\alpha (\Delta
)]\;\;\;\;\;\;\;\;\;\;\;\;\;\left( PBC\right) \;\;\;  \label{RG} \\
\frac{\partial G_\delta (I,J)}{\partial \ln D} &=&\frac 1{2N}{\normalsize %
\sum_{\alpha =-N+1}^{N-1}[G_\alpha (I,J)G_{\delta -\alpha }(I+\alpha
,J-\alpha )-G_{\delta -\alpha }(I+\alpha ,J)G_\alpha (I,J-\delta +\alpha
)]\;\; }  \label{RG2}
\end{eqnarray}
where in $\left( \ref{RG2}\right) \,$the boundary conditions impose that
only couplings with $I,I+\delta ,J$ and $J-\delta $ between 1 and N are
non-zero. The reader may verify that all the RG equations found in \cite
{Solyom79} are reproduced by (\ref{RG}), via the dictionary (\ref{dic}). The
generalized flow invariant is $\sum_\Delta G_0(\Delta )$. It is important to
call attention to the compactness of the formula $\left( \ref{RG}\right) $.
Note that if one takes $\delta $ and $\Delta $ as vectors instead of
scalars, then the couplings can be properly chosen to obtain the RG
equations for models of coupled chains of electrons with spin.

Let us analyze first systems with PBC. For N=2, bosonisation of the model
gives\cite{Giamarchi88}:$\;\mathcal{H}=\mathcal{H}_{TL}+$ $\mathcal{H}%
_{GSG}, $ where $\mathcal{H}_{TL}$ is the Tomonaga-Luttinger hamiltonian$,$
and $\mathcal{H}_{GSG}=1/\left( 2\pi \alpha \right) ^2\int dx\,\left[
-G_1(0)\,\;\cos \left( \sqrt{8}\phi _{-}\right) \;+G_1(1)\,\;\cos \left( 
\sqrt{8}\theta _{-}\,\right) \right] .$ The model consists of the
sine-Gordon model, plus a cosine term in the dual field. For this reason it
is called the \emph{generalized sine-Gordon model (GSG)}. The GSG model is
always massive because it involves dual operators, which thus have inverse
scaling dimensions. There can only be gapless states in hyperplanes of the
phase diagram. A generalization to the N-chain problem leads to the same
conclusion. A simple way of seeing this uses a linearization of the equation
(\ref{RG}), around the possible fixed points. Writing $G_\delta (\Delta
)=G_\delta ^{*}(\Delta )+g_\delta (\Delta )\ $ and linearizing around the
Luttinger Liquid fixed point, we obtain simply: 
\begin{equation}
\partial g_\delta \left( \Delta \right) =1/{2N}\ \left( G_0^{*}\left( \Delta
+\delta \right) +G_0^{*}\left( \Delta -\delta \right) -2G_0^{*}\left( \Delta
\right) \right) g_\delta \left( \Delta \right)  \label{lin}
\end{equation}
We realize that we can always have two symmetrical eigenvalues if N is even:
for $\Delta =\delta =N/2$, $\lambda =1/2N\;[G_0(0)-G_0(N/2)]$ whereas for $%
\Delta =0$, $\delta =N/2$, $\lambda =1/2N$ $[G_0(N/2)-G_0(0)]$. This
corresponds to a competition between two quite different kinds of processes:
in the first case two electrons jump within the same chain, whereas in the
second they exchange chains. Thus we conclude that this competition has a
dramatic effect in destabilizing gapless states in a system where PBC are
taken. An example of systems behaving in this way are the systems of chains
coupled by transverse hopping terms\cite{Yoshioka95}.

Next we investigate the case where no PBC are considered which is the model
with more practical relevance. Here the boundary conditions do not allow
processes like G$_{N/2}\left( N/2\right) $ and so we may think that this
kind of competition is not present. We show next that this competition is
indeed softened, but that it is still present although in a slightly
different fashion. Consider a system of four coupled chains (no PBC). A
simple argument states that if the only non-zero couplings are G$_1\left(
2,2\right) $, G$_1\left( 2,3\right) $,G$_0\left( 2,2\right) $ and G$_0\left(
2,3\right) $ (and all the other couplings that by symmetry are equal to
these), then apart from multiplicative factors, the RG equations have the
same structure as those of the GSG model. Intuitively this arises because we
can look at the first chain as the image of the third and the fourth as the
image of the second, like if we had a model with PBC. Thus we showed that in
this region of the phase diagram a four chain model with open boundary
conditions has the same properties as the GSG model. This subspace is stable
and doesn't really tell us much about the effect of the omitted couplings.
However this provides us with a picture for a kind of competition present
between couplings of the form G$_\delta \left( I,I+\delta \right) $ and G$%
_\delta \left( I,I\right) .$

Consider a model with three spinless chains. We rename the couplings as
shown in figure 1b). Then the RG equations look like: (the factor 1/N is
omitted ) 
\begin{equation}
\begin{tabular}{ll}
$\partial (A+C+D)=\partial (B+2C)=0$ & $\partial (B-2C)=4(E^2-F^2)$ \\ 
$\partial (D+A)=\left( E^2-F^2\right) $ & $\partial E=E(B-2C+G+D)$ \\ 
$\partial G=E^2-F^2+2G(D-A)$ & $\partial F=-F(B-2C+G+A)$%
\end{tabular}
\label{rg3}
\end{equation}
Linearizing around the various fixed points we can verify that the \emph{only%
} stable fixed points are of Luttinger type (E=F=G=0) and lie in the region $%
A\,<\,2C-B\,<\,D$. This seems to hold true for an arbitrary number of chains
as the number of conditions for the eigenvalues of (\ref{lin}) is
proportional to N$^3$ (the order of the number of different {\normalsize $%
G_{\delta \neq 0}(I,J)),$ }whereas the number of relations we can have among
the $G_{\delta =0}(I,J)$ couplings goes as N$^4.$ Thus finding it possible
to have a weak coupling theory for N=2 and 3, it looks likely that a weak
coupling theory exists for any N.{\normalsize \ }

Another point worth noticing is how the E and F couplings do infact compete.
They flow independently of their sign and almost in opposite directions. On
the other hand, the $\delta =0$ processes depend only on the relative
magnitude between the E and F processes, which is a clear manifestation of
their competition. For positive bare couplings the system can reach two
kinds of strong coupling regimes. If F$\rightarrow +\infty $ and $%
E\rightarrow 0,$ then A,B,G$\rightarrow +\infty ,$ D,C$\rightarrow -\infty $
(Regime I)$.$ If E$\rightarrow +\infty $ and $F\rightarrow 0,$ then G,D,B$%
\rightarrow -\infty ,\;A,$C$\rightarrow +\infty $ (Regime II)$.$

We are interested in knowing whether what we found for a system of 3 chains
provides us with a picture for systems with a higher number of chains. Here
we argue that this can indeed be the case for a large class of systems. We
look at systems close to a generalized Luttinger liquid state and for
strictly repulsive interactions. We want to study the influence of small
momenta transfer processes (tangential to the FS) in destabilizing the
massless Luttinger state. For this reason we analyze bare coupling functions
of the general form $G_\delta (I,J)=g(|\delta |)\;f(J-I-\delta )$. The
function $g$ controls the dependence of the couplings on the momentum
transfer. We take it as a positive monotonic (decreasing) function. The
forward scattering models used in studies of 2D bosonization correspond to
the case where $g(|\delta |)$ is a delta function at $\delta =0.$ It's easy
to show that G respects all the necessary symmetries, provided $f$ is an
even function. Here we will present some numerical results. An analytical
study of the equation $\left( \ref{RG2}\right) $ for this class of bare
coupling functions can also be achieved but this is left for future
publication.

In the following numerical results we considered systems of five chains. We
take $g(|\delta |)=\exp \left( -\delta ^2/2\sigma \right) .$ The form of $g$
doesn't seem to play an important role so long as it is even and peaked only
at $\delta =0$. The control parameter $\sigma $ may be seen as anisotropy
dependent: we can expect a curved FS to favor small values of $\sigma.$ For
the function $f$ we take $f\left( x\right) =a_0\left( 1+b_0\;\left(
x/N\right) ^2\right) ,$ like in a Taylor expansion. The coefficients $a_0$
and $b_0\;$are chosen so that $f$ remains positive, as we are interested in
studying the effect of repulsive interactions. In principle it may look
physically more reasonable to take $f$ as a decreasing function. However one
should also remember that $f$ is already a renormalized function where other
effects like those coming from phonons and anisotropy are included.

The results can be summarized as follows. When $f$ is monotonous and
decreasing then no matter how small $\sigma $ is, we fall in a regime where
all chain exchange processes diverge to +$\infty $ and the jump within the
same chain processes decrease (figure 2a ). This corresponds to a
generalization of the regime I (F$\rightarrow +\infty ),$ in the three
chains model. The competition between the two classes of couplings is
obvious. Then we also clearly observe that \emph{all} $G_0\left( n,m\right)
, $ with $n\neq m,$ diverge to $-\infty $ whereas \emph{all }the diagonal
elements $G_0\left( n,n\right) ,$ diverge to +$\infty .$ When $f$ is
monotonous and increasing we can have several regimes. Like in the three
chain model a Luttinger state is possible for small enough $\sigma .$ By
increasing $\sigma $ a massive regime appears similar to regime II ($%
E\rightarrow +\infty )$ of the three chain model (figure 2b ). Now some
processes with jumps within the same chain may also diverge to $-\infty .$
The remaining couplings behave in the same way as in the three chain system.
The competition between each pair of couplings G$_\delta \left( I,I+\delta
\right) $ and G$_\delta \left( I,I\right) ,$ is still present but works in
different ways and magnitudes for each $I$ and $\delta .$ If $\sigma $ is
further increased then we recover regime I, where the jumps within the same
chain processes strongly decrease (figure 2c ). In the three chain model it
is easy to see why this happens: by increasing $\sigma $ we increase the
coupling G, which is a chain exchange process. Thus this class of processes
is favoured and manages to overcome the jump within the same chain
processes. This analysis seems to apply for any finite N. We would like to
stress that the three chain model seems to reproduce much of the physics of
a more general N chain model.

Finally, we consider a 2D electron gas, where flat portions are introduced
in a circular FS (figure 3). We take N patches in the nested regions, and M
in the curved portions, with N/M$\ll 1$. Due to the geometry of the FS, it
is more appropriate to parametrize the various patches in terms of their
angular positions. This is achieved through the substitution $G_\delta
\left( I,J\right) \rightarrow \widetilde{G}_\delta \left( I,N+M+1-J\right) $%
. The symmetries on the square restrict the number of independent couplings:
0$\leq \delta \leq N+M,$ 1$\leq I,J\leq N+M.$ Using Shankar's arguments\cite
{Shankar94}, we neglect any zero sound contributions to a vertex involving
electrons on the curved portions. This corresponds to introducing a factor $%
\delta _{I\in \phi }\delta _{I+\delta \in \phi }$ on the second term of $%
\left( \ref{RG2}\right) $ ($\phi $ stands for the nested region). Also, the
only couplings with electrons on these regions have the BCS form $\widetilde{G}%
_\delta \left( I,I\right) $. With these modifications, the equation $\left( 
\ref{RG2}\right) $ is still valid, if we regard the couplings as already
renormalized by the finite 2D density of states. We linearized the RG equations around the
Luttinger fixed point for a system with N=3, M=0 and concluded that the
Luttinger regime remains stable if a further condition on the couplings B,D$>$0, 
is verified. The other strong coupling regimes do also
exist. If the couplings involving scatterings not restricted to a nested
region are sufficiently strong, then the regime II leads to a new regime,
where the E coupling goes to -$\infty $ (figure 3). This state seems to be a
good candidate for a superconductive state. In fact, in this regime F$%
\rightarrow 0,$ so that the main corrections to the vertices come from the
Cooper channel, and are due to effective attractive interactions. This will
be studied in detail elsewhere\cite{Vistulo96}.

We should point that the possibility of generating ordered phases due to
nesting is not a new concept: it remains the main explanation for itinerant
antiferromagnetism,it has been suggested 
to explain the HTCSC, and it may explain the
stability of some CDW phases recently observed\cite{Dzyaloshinskii72}.

We acknowledge discussions with A.M.Tsvelik. FVA thanks discussions with R.
Chitra, L. Laloux, B. Delamotte, J.M.Robin and financial support from EEC
under the grant ERBICHBICT941242.

\newpage
{\textbf Figures Captions}\\

\epsfig{file=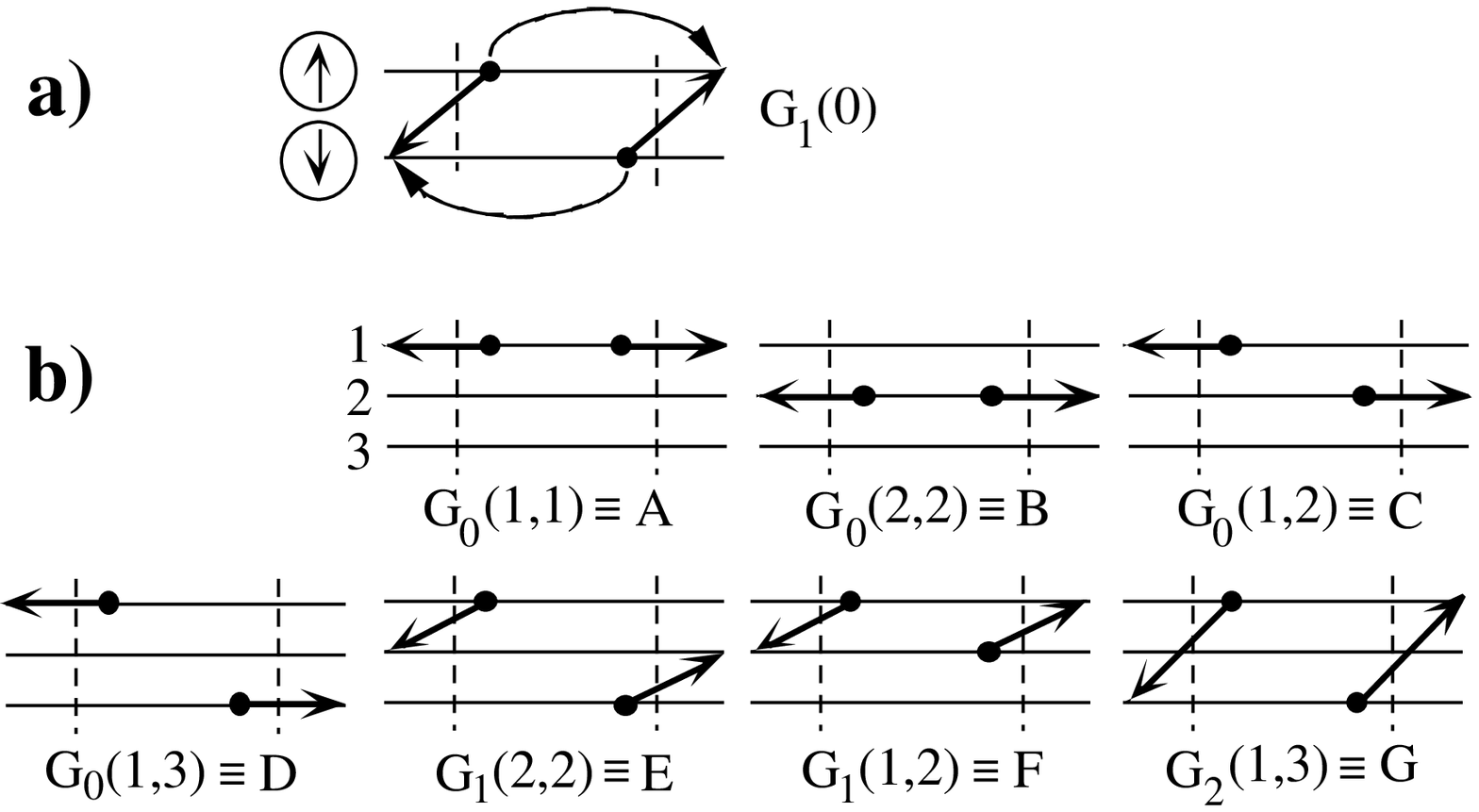,width=12cm}\\
Figure 1. a) One-dimensional interacting electrons with spin can be seen as
two spinless chains in interaction. The backscattering process (dashed
arrows), where two electrons change from one side of the FS to the other, is
now seen as a spin-flip process (solid arrows). In b) all scattering
processes involved in the three chain model with no PBC are shown.

\epsfig{file=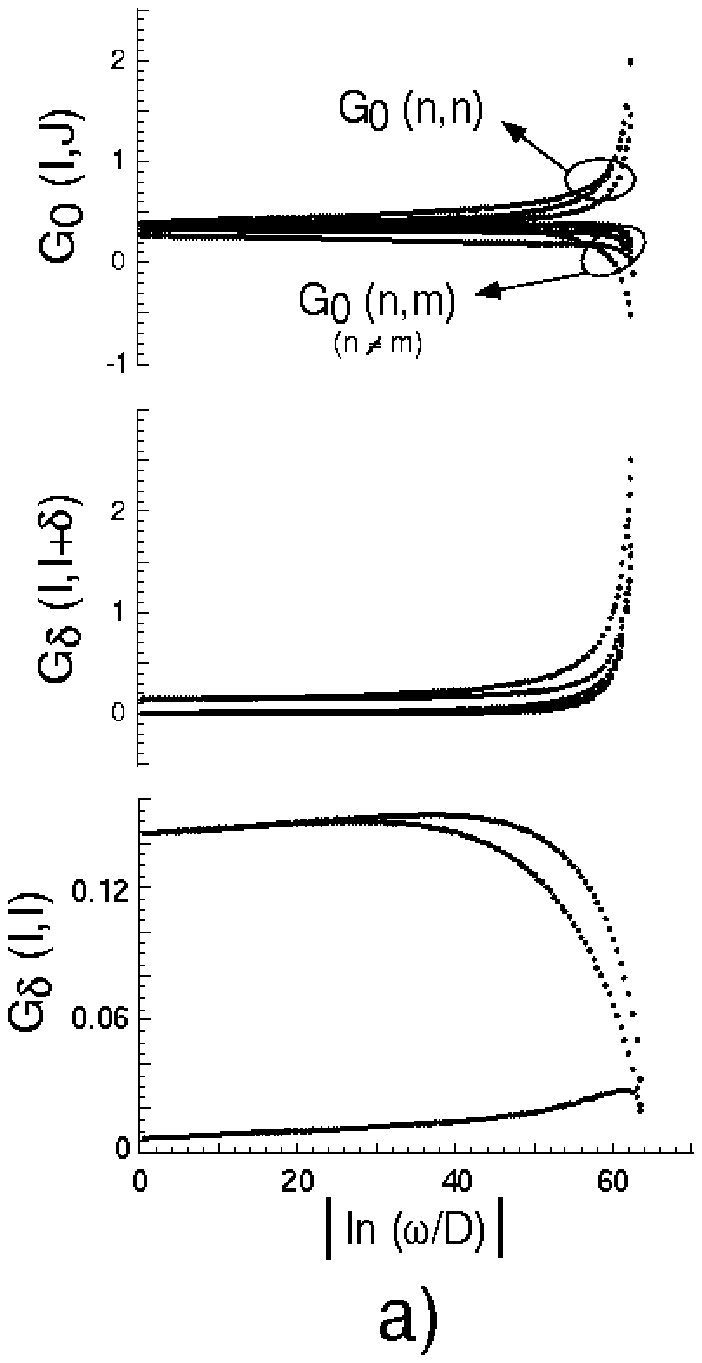,width=7cm}
\epsfig{file=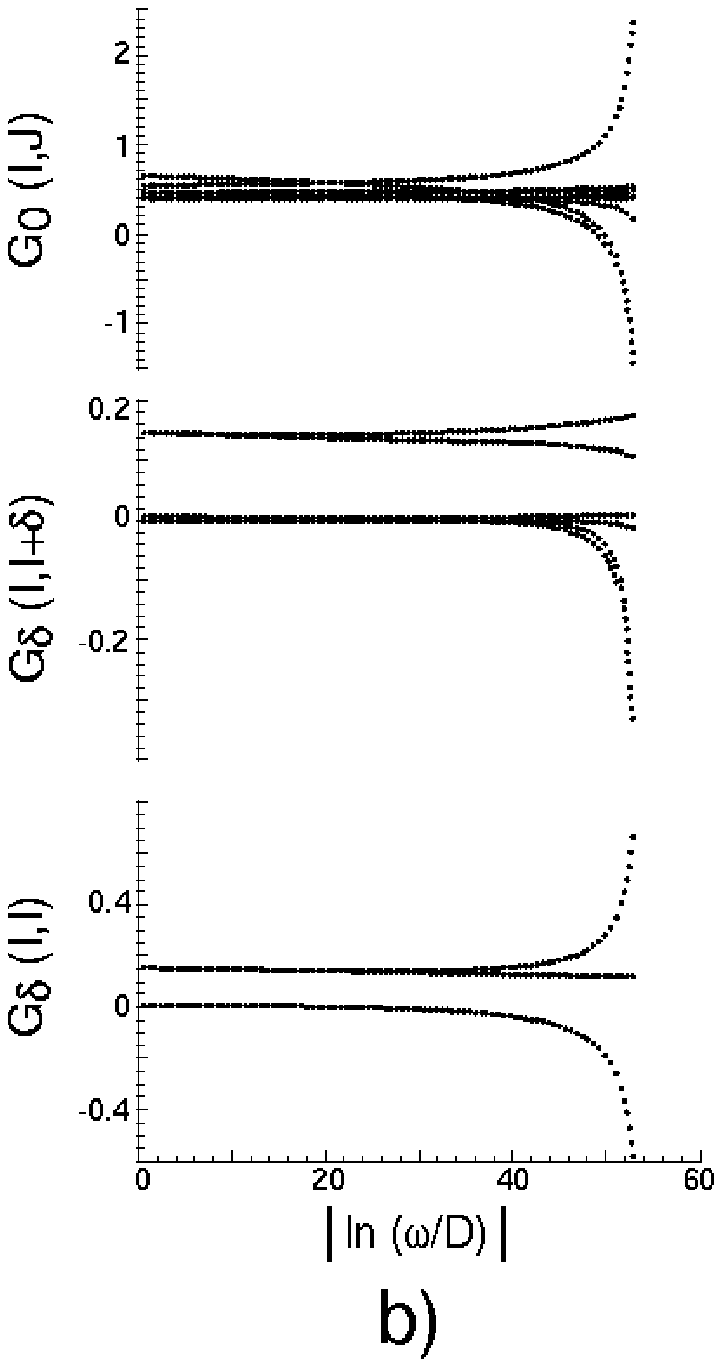,width=8cm}
\epsfig{file=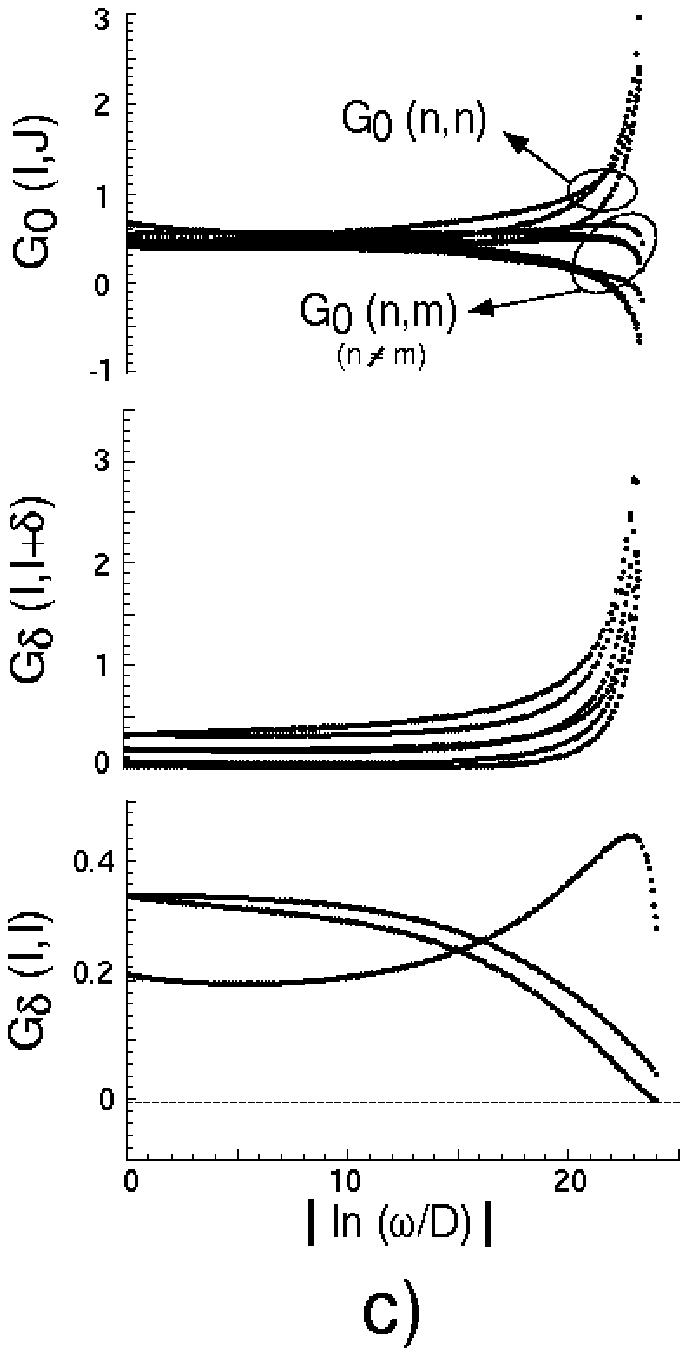,width=7cm}

Figure 2. 
Strong coupling regimes: a) $f\left( x\right) =0.4$ $\left(
1-1/2\;\left( x/N\right) ^2\right) $ and $\sigma =0.5;\;$\\b) $f\left(
x\right) =0.4$ $\left( 1+\left( x/N\right) ^2\right) $ and $\sigma =0.5;$ c)
same $f\left( x\right) $ but $\sigma =2.5$%

\epsfig{file=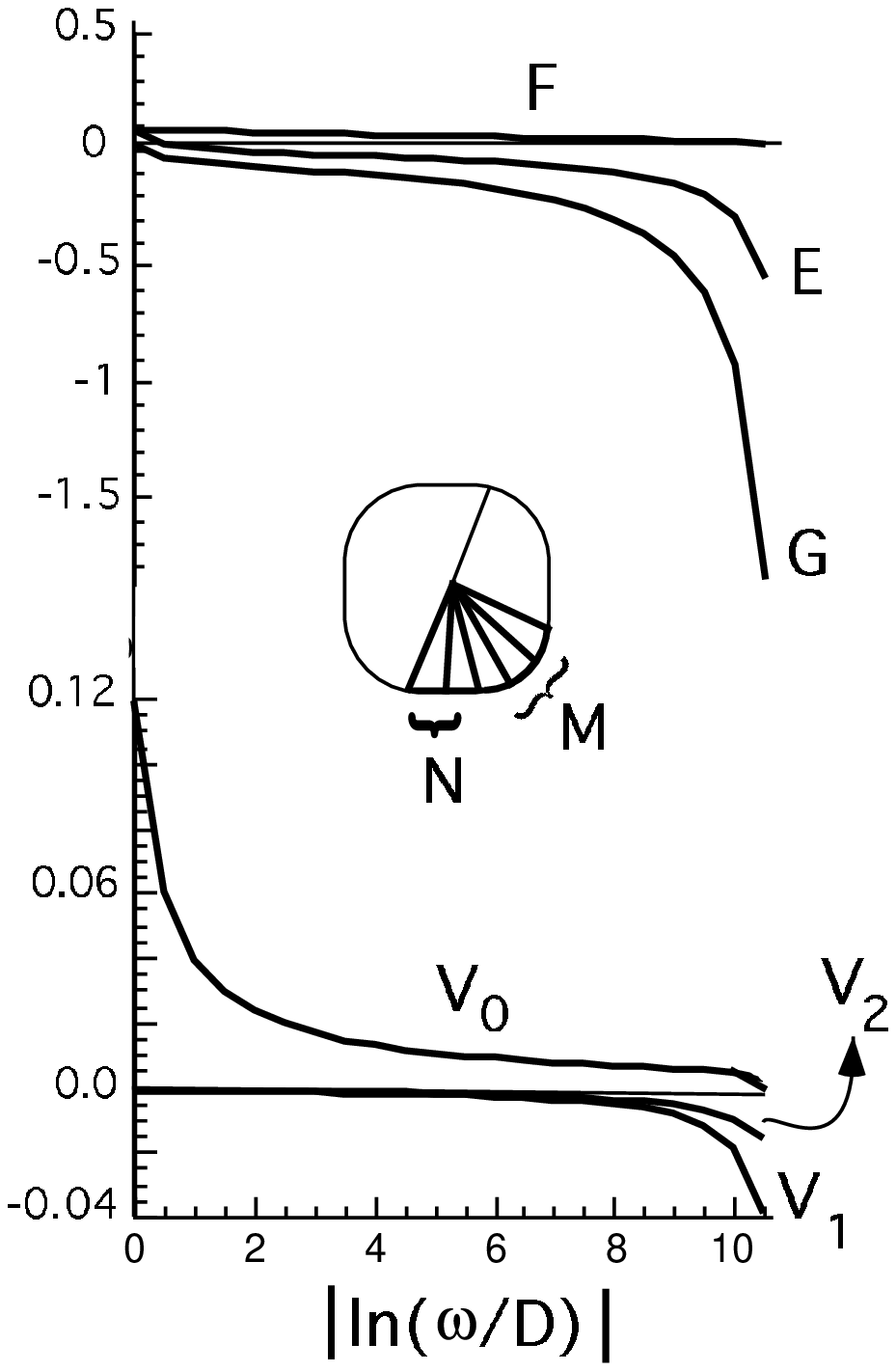,width=10cm}

Figure 3. 
For a system with N=3, M=14, bare couplings A=B=C=D=0.1, E=F=0.08,
G=0.02 and the remaining couplings equal to 0.12, the Fourier modes V$_l$ of 
$\sum_I\widetilde{G}_\delta \left( I,I\right) $ diverge to -$\infty ,$
whereas F$\rightarrow 0.$

\end{document}